\documentstyle[11pt,epsf]{article}
\def\beq{\begin{eqnarray}}
\def\eeq{\end{eqnarray}}
\def\bea{\begin{eqnarray*}}
\def\eea{\end{eqnarray*}}

\def\centeron#1#2{{\setbox0=\hbox{#1}\setbox1=\hbox{#2}\ifdim
\wd1>\wd0\kern.5\wd1\kern-.5\wd0\fi
\copy0\kern-.5\wd0\kern-.5\wd1\copy1\ifdim\wd0>\wd1
\kern.5\wd0\kern-.5\wd1\fi}}
\def\ltap{\;\centeron{\raise.35ex\hbox{$<$}}{\lower.65ex\hbox{$\sim$}}\;}
\def\gtap{\;\centeron{\raise.35ex\hbox{$>$}}{\lower.65ex\hbox{$\sim$}}\;}

\def\singleandthirdspaced{\baselineskip=\normalbaselineskip\multiply
    \baselineskip by 130\divide\baselineskip by 100}

\newcommand{\newc}{\newcommand}
\newc{\qbar}{{\overline q}}
\newc{\Kahler}{K\"ahler }
\newc{\deltaGS}{\delta_{\rm GS}}
\begin{document}
\begin{titlepage}
\begin{flushright}
{\large hep-th/yymmnnn \\ SCIPP 07/18\\
}
\end{flushright}

\vskip 1.2cm

\begin{center}

{\LARGE\bf Dynamical Supersymmetry Breaking and Low Energy Gauge Mediation}

\vskip 1.4cm

{\large Michael Dine and John D. Mason}
\\
{\it Santa Cruz Institute for Particle Physics and
\\ Department of Physics, University of California,
    Santa Cruz CA 95064  } \\
\vskip 4pt

\vskip 4pt

\vskip 1.5cm

\begin{abstract}

Dynamical breaking of supersymmetry was long thought to be
an exceptional phenomenon, but recent developments have altered
this view. A question of great interest in the current framework
is the value of the underlying scale
of supersymmetry breaking.
The ``little hierarchy" problem suggests that supersymmetry should be broken at low
energies.    Within one class of models, low energy
breaking be achieved as a consequence of symmetries, without requiring
odd coincidences.    The low energy theories
are distinguished by the presence or absence of
$R$ symmetries; in either case, and especially the latter
one often finds modifications of the minimal gauge-mediated spectrum
which can further ameliorate problems of fine tuning.  Various natural
mechanisms exist to solve the $\mu$ problem in this framework.
\end{abstract}

\end{center}

\vskip 1.0 cm

\end{titlepage}
\setcounter{footnote}{0} \setcounter{page}{2}
\setcounter{section}{0} \setcounter{subsection}{0}
\setcounter{subsubsection}{0}

\singleandthirdspaced

\section{Introduction}

If supersymmetry is relevant to low energy physics, supersymmetry
breaking is probably dynamical; indeed,
the possibility of dynamical breaking (DSB) is a principal reason to consider
low energy supersymmetry\cite{wittendsb}.

In the past, this was a complicated topic.  Models with gravity mediation
were problematic; it was hard to suppress
flavor changing neutral currents and to obtain a suitable
gaugino spectrum.   Models
with low energy gauge mediation did not suffer
from these problems, and lead to certain generic
predictions.  However, these early models were so complicated as to
appear contrived.  It was challenging to generate
$\mu$ and $B_\mu$ terms for Higgs fields,
and the theories often exhibited approximate $R$ symmetries.
Moreover, as the limits on the Higgs mass and the masses of
various superparticles have improved, problems of fine tuning have become
severe.

Recent developments in dynamical supersymmetry breaking\cite{iss,dfs,shih,dinemason,
aharonykachrusilverstein,seibergreview}
reopen these issues.
As we will discuss, they permit the construction of simpler models.  In this
richer set of theories, the predictions of the simplest models of gauge
mediation (Minimal Gauge Mediation, or MGM)
do not necessarily hold\cite{shihnew,bankskorneel}.
This has the potential to reduce fine tuning.
Lowering the scale of supersymmetry breaking to energies of order $10$'s to $100$'s
of TeV also reduces the amount of fine tuning.  While challenging in earlier
theories, this can readily be achieved with the present understanding.
The issues of $\mu$ and $B_\mu$, and gaugino masses, take
on a different character in these new theories.

This paper provides a perspective on these developments.  Our focus
will be on models in which supersymmetry is broken
at the lowest scales commonly contemplated for gauge mediation,
of order $100$ TeV.  This is a quite restrictive requirement.  The ISS
models, for example, involve two mass scales, one the scale of the new
gauge interactions, $\Lambda_N$, the other the scale of the masses of
the hidden-sector matter fields, $m_f$.  The scale of supersymmetry
breaking is suppressed relative to $\Lambda$ by a power of $m_f/\Lambda$.
So unless $m_f$ is comparable to $\Lambda$, the scales $\Lambda$ and $m_f$
must be much larger than $100$ TeV. Similar issues arise in many other models
(in the "Pentagon" models of Banks and collaborators\cite{pentagon}, this coincidence is argued
to arise from an underlying principle of theories of quantum gravity).

More generally,
we will see that:
\begin{enumerate}
\item  In many of the new models, achieving DSB at a low scale
still requires surprising
coincidences or baroque constructions.  But in a
broad class of models (the ``retrofitted" models of \cite{dfs} and
generalizations\cite{aharonykachrusilverstein}), one can achieve
low energy breaking in a rather simple way, consistent with conventional
notions of naturalness.  
\item  Many of the new models possess,
at low energies, approximate $U(1)_R$ symmetries.  These must be spontaneously
broken.  This requires the presence of gauge symmetries\cite{dinemason,iss2},
or fields $R$ charge different than $0$ or $2$\cite{shih}.  With spontaneous breaking,
suppression of CP violation is sometimes automatic\cite{shihnew}.  However,
in these cases, it is necessary to suppress dangerous one loop contributions
to squark and slepton masses.  
\item  Explicit breaking of $R$ symmetry can be obtained by retrofitting.
In the retrofitted models, the $R$ symmetry breaking scale can naturally
be of order the supersymmetry breaking scale, allowing a low scale for 
supersymmetry breaking.
To obtain metastable
breaking naturally, a small parameter is required, which can arise
through a Froggatt-Nielsen (FN) type mechanism; the underlying scale
of supersymmetry breaking is then two orders of magnitude or more larger
than the messenger scale.
With explicit breaking, suppression of electric dipole
moments
is not automatic; additional features are required to suppress CP-violating
phases.  
\item  A suitable $\mu$ term can be generated in the retrofitted models,
consistent with conventional notions of naturalness, but we will see that
additional degrees of freedom are required.  We will consider two
classes of models, one in which $\tan(\beta)$ is of order one\cite{giudice,
giudicetwo,dfs}; another
in which $\tan(\beta)$ is automatically large.  
\end{enumerate}

In the next section, we will explain why the little hierarchy problem, in the framework
of gauge mediation, suggests that the supersymmetry breaking scale should be low,
and the superparticle spectrum {\it compressed} or {\it squashed}\cite{agashegraesser,
dineminnesota,shihnew,bankskorneel}.  In section 3,
we will review the classification developed in \cite{seibergreview} of theories exhibiting
supersymmetry breaking, with a view
to low energy gauge mediation.  In most, significant additional structure is required
in order to mediate supersymmetry breaking, especially if the
supersymmetry breaking is to arise at a low scale.  Additional
scales must be introduced, and the scale of supersymmetry
breaking at low energies is typically a ratio of various
more microscopic mass scales. 
  The situation is different
in the class of models developed
in\cite{dfs,aharonykachrusilverstein}.  Here one begins
with a (generalized) O'Raifeartaigh model, and understands the coefficients of
relevant operators as dynamically generated by physics at some much higher energy
scale.  In section 4, we will discuss (elaborating ideas in \cite{dinemason})
how coefficients
of relevant operators of different dimensions can all be given
by powers of the same mass scale;
this is necessary if the 
supersymmetry breaking scale is to be low.  We explain why these
models frequently possess continuous R symmetries
at low energies, and discuss both spontaneous and explicit
breaking of these symmetries.  In the resulting
models, we explain why the spectrum need not be that
of minimal gauge mediation (MGM), and can be compressed.   Finally,
we discuss mechanisms for generating suitable $\mu$ and $B_\mu$.
A proposal of Giudice and Rattazi is readily implemented in this framework
and can lead to moderate $\tan \beta$; another mechanism, closer
to the spirit of the retrofitted models, leads to a large
value of $\tan(\beta)$.

\section{Implications of the Little Hierarchy}
\label{littlehierarchy}

At present, the MSSM appears to be fine-tuned at the one
percent level.  While there are some regions of the parameter
space where the difficulties may not be so serious\cite{focuspoint}, and
additional degrees of freedom may ameliorate the problem,
the tuning is even more
severe in gauge mediated models.  In general supersymmetric
models, there are two sources of difficulty.
First, in order to enhance the Higgs quartic coupling so as to be compatible
with the LEP bound, one requires $\tilde m_t > 800$ GeV, and/or substantial
stop mixing.  This, in turn, implies a large correction to the Higgs mass from
stop loops:
\beq
\delta m_H^2 = -{12 y_t^2\over 16 \pi^2} m_{\tilde t}^2 \ln(M/m_{\tilde t}).
\eeq
If the cutoff, $M$, is of order the Planck scale, and $m_{\tilde t} \sim 800$ GeV,
then
\beq
\delta m_H^2/m_Z^2 \approx 189
\eeq
which implies a severe fine tuning.  This problem can be ameliorated if $m_{\tilde t}$ is
smaller.  This requires the presence of additional dynamics beyond the MSSM\cite{bmssm}.If, for
example, $\tilde m_t =300$ GeV, $\delta m_H^2/m_Z^2 \approx 27$, suggesting a fine tuning of
order $4\%$.  

The situation is perhaps worse in MGM models, since even if
there are extra degrees of freedom which can explain the Higgs mass, the
mass of the stop is large.  In MGM, the messengers
lie in $5$ and $\bar 5$ representations of the usual
$SU(5)$, and couple to a gauge singlet, $S$,
with $\langle S \rangle = s + \theta^2 F_S$.  Then
\beq
\widetilde m^2 ={2 \Lambda^2} 
\left[
C_3\left({\alpha_3 \over 4 \pi}\right)^2
+C_2\left({\alpha_2\over 4 \pi}\right)^2
+{5 \over 3}{\left(Y\over2\right)^2}
\left({\alpha_1\over 4 \pi}\right)^2\right],
\label{scalarsmgm}
\eeq
where $\Lambda = F_S/s$.
$C_3 = 4/3$ for color triplets and zero for singlets,
$C_2= 3/4$ for
weak doublets and zero for singlets.
This formula predicts definite ratios of squark,
slepton and gaugino masses.  Coupled with the current limits on the lightest
sleptons (approximately $100$ GeV), it implies that  
slepton doublets have masses greater than $215$ GeV, while squark masses are larger than
$715$ GeV.  The cutoff, $M$, in gauge-mediated models, is of order
the messenger scale.  Assuming a value of this scale of order
the GUT scale, as in many models,
\beq
\delta m_h^2/M_Z^2 \approx 130;
\eeq
if $M= 10^2$ TeV, $130$ is reduced to $21$, suggestive of a $5\%$ fine tuning. 
  So from the perspective of fine tuning, a low value of this
scale is preferable.  If the stop quark mass were not much above the current
limit (say $300$ GeV), and $M$ were of this order, the fine tuning
would be insignificant.

The challenge for gauge-mediated model building, then, is to obtain slepton singlets not
much above the current experimental bound, doublet masses not much larger, and triplet
masses not much above $300$ GeV, while breaking supersymmetry
at a low scale.  This compression of the spectrum requires
a structure different from that of minimal gauge mediation, which 
we will refer to as ``General Gauge Mediation", or GGM.  Examples
of such structures have appeared in the literature (see, e.g.,
\cite{shih}).\footnote{Such compression of the spectrum has been
considered by many; N. Arkani-Hamed, N. Seiberg, and S. Thomas
have all mentioned this possibility to us.  Indeed, even
the structure of the earliest gauge-mediated model\cite{dinefischler}
is not that of MGM.}  This can be achieved in a model
with multiple gauge singlets, $S_i$.  For example, with a single
$5$ and $\bar 5$ of messengers, with $5 = q + \ell$, $\bar 5 = \bar q + \bar \ell$,
\beq
W= \lambda_i S_i \bar q q + \gamma_i S_i \bar \ell  \ell
\label{lambdadef}
\eeq
where
\beq
\langle S_i \rangle = s_i + \theta^2 F_i,
\eeq
we can define:
\beq
\Lambda_q = {\lambda_i F_i \over \lambda_i s_i}~~~
\Lambda_\ell = {\gamma_i F_i \over \gamma_i s_i}
\eeq
The masses of the gluinos are given by
\beq
m_{\lambda} = {1 \over 2} {\alpha_3 \over 4 \pi} \Lambda_q~~~~
m_w = {1 \over 2} {\alpha_2 \over 4 \pi} \Lambda_{\ell}~~~~
m_b = {1 \over 2} {\alpha_1 \over 4 \pi} \left [2/3 \Lambda_q + 
\Lambda_\ell \right ].
\label{gluinoformula}
\eeq
Similarly, for the squark and slepton masses we have: 
\beq
\widetilde m^2 =2 
\left[
C_3\left({\alpha_3 \over 4 \pi}\right)^2 \Lambda_q^2
+C_2\left({\alpha_2\over 4 \pi}\right)^2 \Lambda_\ell^2
+{\left({Y\over2}\right)^2}
\left({\alpha_1\over 4 \pi}\right)^2 ({2 \over 3} \Lambda_q^2 + \Lambda_\ell^2) \right],
\label{scalarsemgm}
\eeq

The breakdown of the simple relation (``unification relation") for gaugino masses
is potentially problematic for electric dipole moments, since the phases of
the various gaugino masses are no longer identical.  As we will discuss shortly,
some suppression of CP violating phases is almost certainly required
in such a situation.

Recently, Cheung et al\cite{shihnew} have discussed
models with multiple messengers and $R$ symmetries,
with a single field, $S$, coupling to the messengers.
In this case, they show that the squark and slepton masses
can be compressed, while, perhaps surprisingly, the
gaugino masses are still unified.  So this class of models
has less difficulty with questions of CP violation, then
general gauge-mediated models.  We will discuss later how
these models can emerge naturally with low scale supersymmetry
breaking.  There is another potential difficulty with these models, however.
In models such as that of eqn. \ref{lambdadef}, there is an a approximate
left-right symmetry under which the hypercharge $D$ term is odd.  But in the
models with multiple messengers, this is not automatically the case.  As a result,
there can be one loop contributions to squark and slepton masses, of both positive
and negative sign\cite{dinefischler}.  On the other hand, as we will explain elsewhere,
such an approximate {\it messenger parity} symmetry can arise as an accident even in
more complicated models.

In retrofitted models with explicit breaking of the R symmetry,
the formulas of minimal gauge mediation also do not necessarily hold.
As in the case of multiple singlets, some suppression of
CP violating phases is generally required to suppress electric
dipole moments.

\section{Survey of DSB Models}

Recently, there has been an appreciation that metastable supersymmetry breaking 
is a rather generic phenomenon in supersymmetric theories, occurring even in theories
which are non-chiral and posses non-vanishing Witten index.
It has been noted that these models open many new possibilities for
high energy supersymmetry breaking\cite{murayamaiss,aharonyseiberg,dfs}.
Here our focus is on the possibility of using such models for low energy breaking.
Ref. \cite{seibergreview}
distinguishes four classes of theories which exhibit dynamical supersymmetry
breaking (DSB):
\begin{enumerate}
\item  The $3-2$ class:  these were the earliest theories of DSB which were well
understood.  These models are distinguished by a chiral
structure, an absence of pseudomoduli (flat directions)
and $R$ symmetries, at the level of non-renormalizable operators (more generally,
operators up to a given dimension).  In known examples, the $R$ symmetry is spontaneously
broken.  The model with gauge group $SU(3)\times SU(2)$ and fields with the quantum
numbers of a single generation, is the simplest example in this class.  If one writes
general, higher dimension operators in these theories, one finds supersymmetric minima
at large fields.  So one expects that even in these models, the vacuum is (highly)
metastable.
\item  The ISS class:  The simplest model in this class is the $SU(N)$ gauge theory
with $3/2 N > N_f > N+1$ studied in \cite{iss}.  The models possess,
in their simplest versions, distinct  
features:  they are non-chiral, they possess no classical flat directions, and
they have non-vanishing Witten index.  Known examples possess parameters with dimensions
of mass.  These masses can be given a dynamical explanation if further gauge interactions
are introduced (as in the retrofitted models below)\cite{
murayamaiss,aharonyseiberg}.  In simple examples, the susy-breaking
vacua of these models exhibit an unbroken $R$ symmetry.  This feature is probably
not fundamental.  
It is also possible to break the symmetries of the models explicitly, again
through retrofitting.  In simple models, achieving a low scale of supersymmetry
breaking requires coincidences, as we will explain.
\item  The ITIY class:  Models in this class have been constructed by coupling
gauge singlets to a theory with a quantum moduli space\cite{itiy}.  More generally, these theories
are characterized by coupling of singlets to a theory with flat directions (and, in
particular, no mass gap); the resulting construction has a pseudomoduli space of 
vacua.  The models possess a continuous $R$ symmetry at the level of low dimension
operators.  This symmetry may be spontaneously broken in the presence
of additional weak gauge interactions.  Recently, Ibe and
Kitano have speculated on the existence of metastable
$R$-breaking minima in the strong coupling region\cite{kitano}.
If this assumption is correct, this opens additional model-building strategies.
Explicit breaking, as explained below,
may require coincidences.
\item  The retrofitted class:  These are (generalized) O'Raifeartaigh models,
in which some or all dimensionful parameters arise as a consequence of non-perturbative
dynamics at a high energy scale.  In the original proposal, the non-perturbative dynamics
was associated with some new strong interactions.  However, as pointed out in
\cite{aharonykachrusilverstein}, string-scale instantons (and presumably other very
high energy effects) could generate such couplings.
\end{enumerate}

In the following subsections, we consider the first three classes in turn as models of
low energy DSB.   We address each of the phenomenological issues we enumerated
above: the scales of new dynamics,
$R$ symmetry breaking, and the origin of the $\mu$ and $B_\mu$ terms.  The
next section is devoted to the retrofitted models.

\subsection{The $3-2$ Class}

Many models in this class admit the possibility of large global symmetry groups.
One can try to identify a subgroup of this global symmetry group with the symmetry
of the Standard Model.   This provides a model of {\it direct mediation}:  there is
no limit of such a theory in which supersymmetry remains broken and the messengers
decouple.  Such a theory would yield a spectrum of gauginos, squarks and sleptons
like that of gauge mediation, and in the simplest cases like that of minimal
gauge mediation.  The scale of supersymmetry breaking
can well be very low.  There are, however, two difficulties with such a scheme:
\begin{enumerate}
\item  Asymptotic freedom:  because the underlying strong groups are typically
quite large, asymptotic freedom of the various gauge groups is rapidly lost.
\item  Higgs:  unless the Higgs fields somehow emerge from the strong dynamics
(along the lines of little higgs theories), they do not couple through renormalizable
operators to the fields which break supersymmetry.  As a result, $\mu$ and $B_\mu$
are protected by approximate symmetries.  It is necessary to add additional
degrees of freedom, such as singlets or additional $U(1)$ gauge fields, which
couple to the susy breaking sector.
\end{enumerate}

An alternative approach is to take for messengers additional degrees of
freedom with
Standard Model quantum numbers.  It is then necessary that there be
additional degrees of freedom which couple both to messengers
and to the supersymmetry breaking sector.  Workable models of this type have been
constructed, but they are typically quite complicated, and the underlying SUSY-breaking
sector tends to be at a quite high energy scale, because
of the extra loop factors involved\cite{dinenelson}.

\subsection{The ISS Class}

The ISS models greatly broaden the class of susy-breaking theories.  However,
in models constructed to date, in addition to the scale, $\Lambda$,
of the strong dynamics, another scale is required ($m$, the quark mass in the
simplest models).  To achieve low energy supersymmetry breaking, the scales
$\Lambda$ and $m$ must be similar.  Even if $m$ arises dynamically,
a rather remarkable coincidence is required.

Putting this concern aside, one can explore different approaches to model building.
The ISS models often admit large global symmetries, as in the $3-2$ case.  Constructing
models of direct mediation runs into difficulties with asymptotic freedom and
with $\mu$ and $B_\mu$ terms, as above.  The simplest models have the further
difficulty that they possess approximate unbroken R symmetries.  To date,
most model building with these theories has involved the
introduction of $R$ symmetry breaking dynamics at very high scales, with $\mu$
and $B_\mu$ terms generated by similar dynamics. The scale of supersymmetry
breaking in these models is quite large, typically intermediate between the 100 TeV
scale and $10^{8}$ TeV.  Seiberg and Shih\cite{iss2} and Shih\cite{shih}
have explored alternative possibilities for breaking the $R$ symmetry.
Banks and collaborators
have followed a different approach.  They assume that in a model with
non-calculable strong dynamics (the ``pentagon model"), the $R$ symmetry is broken.  The required
coincidence of scales is assumed to originate from principles of an underlying
theory of gravity (``cosmological supersymmetry breaking").  The pentagon model
has five additional vector-like $5$ and $\bar 5$'s in the sense of ordinary $SU(5)$,
so the unified coupling is perhaps just barely small enough to account
for perturbative
unification.

\subsection{The ITIY Class}

Many of the comments of the previous sections apply to the ITIY models.  Again, it is 
possible to obtain large global symmetries which can be gauged, but this
leads to difficulties with asymptotic freedom.  Similar to the ISS case, the simplest
models do not break $R$ symmetry at the scale of the underlying
strong dynamics, and again, most model building along these
lines\cite{murayamaitiy} assumes $R$ symmetry and supersymmetry breaking at 
a very high energy scale.  Recently, conjecturing that there is a metastable vacuum
in the strongly coupled region, Ito and Kitano\cite{kitano} have constructed a
potentially viable model with low scale
messengers.  However
because there are five additional vector-like fields, unification
may be problematic, as in other instances we have discussed.

\section{Model Building in Retrofitted Models}

What has been called retrofitting is a simple realization of Witten's original
vision for dynamical supersymmetry breaking\cite{wittendsb}:  an exponentially
small coefficient of a relevant
operator is generated by dynamical effects, precipitating supersymmetry
breaking.

The basic ideas of retrofitted models can be illustrated with the simplest O'Raifeartaigh
model:
\beq
W = Z(A^2 - \mu^2) + m YA.
\label{ormodel}
\eeq
Here we have not explicitly indicated a dimensionless number (the coefficient of $Z A^2$)
in the superpotential; we will frequently do this to avoid introducing names for parameters
not germane to our main arguments.
The idea is to replace the parameters $\mu$ and/or $M$ with dynamically
generated scales.  One simple way to do this is to replace $\mu^2$ in the
superpotential by $W_\alpha^2/M$, where $W_\alpha$ is the field strength
of some new, strongly interacting group (which does not break susy), i.e. to
replace $W$ by:
\beq
\int d^2 \theta \left (Z (A^2 - {W_\alpha^2 \over M}  ) + m YA \right ).
\label{retro}
\eeq
Then $\mu^2 = \Lambda^3/M$, where $\Lambda$ is the scale 
of the new gauge group.

An even simpler version of this idea, in the sense that the number of degrees
of freedom at low energies is smaller,
is provided by the work of \cite{aharonykachrusilverstein}.
These authors exhibit string theory constructions where parameters such as
$\mu^2$ or $m$ are generated by string scale instantons.

For low scale supersymmetry breaking, one needs $\mu \sim m$.
As explained in \cite{dinemason}, this can arise if the underlying theory
possesses a discrete $R$ symmetry, with the dynamical
source of the parameters $\mu$ and $m$
transforming under the symmetry.
The idea that the dimensionful parameter, $\Lambda$, arising in
strong dynamics or instanton computations 
can transform under a discrete (gauge) symmetry is a familiar one.
Consider, for example,
\beq
\int d^2 \theta (Z (A^2 - {W_\alpha^4 \over M^4}) + {W_\alpha^2 \over M^2} YA.
\label{ww}
\eeq 
With $W_\alpha^2 \rightarrow e^{2 \pi i k/N} W_\alpha^2$,
$A \rightarrow e^{2 \pi i k/N} A$, and $Y$ and $Z$ are neutral (it is also necessary
to impose a $Z_2$, under which $A$ and $Y$ are odd), this is the most general
structure consistent with symmetries.

\subsection{R Symmetries:  Spontaneous Breaking}

In eqn. \ref{ww}, we have exhibited only the lowest dimension terms (relevant and marginal
operators constructed from $A,Z,Y$).  At this level, the theory has an accidental,
continuous $R$ symmetry.  The appearance of such symmetries is typical; in the next
section, we will explore ways to avoid them.  
In building realistic models, with messengers, (spontaneous)
breaking of the R symmetry is crucial.
It is well known that in this simple model, there is no vacuum
close to the origin which breaks the $R$ symmetry.
As shown by Shih\cite{shih}, this result is general for theories without gauge interactions,
and with the restriction that all $R$ charges are $0$ or $2$; with more general
$R$ charges, spontaneous breaking can occur.  With gauge interactions,
or with fields with negative $R$ charge, the symmetry may be spontaneously broken\cite{dinemason,
iss2}.
In \cite{dinemason}, for example, a model was constructed with additional
gauge interactions, a single neutral field, $Z$, and charged fields $Z^\pm, \phi^\pm$.
The superpotential was taken to be:
\beq
W= M(Z^+ \phi^- + Z^- \phi^+) + \lambda Z^0(\phi^+ \phi^- -\mu^2).
\label{dm}
\eeq
Here $Z^{\pm},Z^0$, at tree level, have non-zero $F$-components; at one loop, for a range
of parameters, they have non-zero scalar components as well, breaking the $R$ symmetry.
$Z^0$, for example, can then be coupled to messengers neutral under the $U(1)$.  As in eqn.
\ref{ww}, the scales $\mu$ and $M$ can naturally be of the same order, so
low energy breaking can be natural\cite{dinemason}.

In simple models, with only
a small number of susy breaking fields and
a single pair of messengers, one obtains the spectrum of minimal gauge mediation.
With more fields, the predictions may be modified (and one might obtain a compressed
spectrum).  As we will explain later, there are a variety of
ways to obtain a $\mu$ term with a suitably small $B_\mu$ in these circumstances.
While natural,  $R$ symmetry breaking
requires that the new gauge  coupling to be similar in value
to the Yukawa coupling\cite{iss2}.  So it is interesting
to consider mechanisms which might give rise to an explicit breaking of the $R$ symmetry
(i.e. for which there is no accidental $R$ symmetry in the low energy theory).

As we will discuss in the next section, it is possible, in the retrofitted framework,
to formulate theories without $R$ symmetries at low energies.  But first, we note that
Cheung et. al.\cite{shihnew} have considered a larger
class of theories with $R$ symmetries at low energies.  They consider
theories with a single supersymmetry-breaking field, $X$, coupled to multiple
messengers, with the quantum numbers of $5$ and $\bar 5$'s of $SU(5)$, $5_i$ and
$\bar 5_i$.  They take the general superpotential
\beq
W = \lambda_{ij} X 5_i \bar 5_j + m_{ij} 5_i \bar 5_j
\eeq
and assume that the couplings preserve a continuous $R$ symmetry.  While the couplings
and fields we have written above are shorthand for couplings to fields $q_i, \bar q_i$
and $\ell_i, \bar \ell_i$, and the couplings to triplets and doublets need not be identical,
it is important in their analysis that the $R$ charges of doublets and triplets {\it are}
the same.  They then show, as we indicated before, that the formulas of MGM for the scalar
masses are modified, but that the gaugino masses do obey unification relations.  This
has promise for ameliorating the little hierarchy problem while possibly solving
the problems of CP violation in supersymmetric theories. It is straightforward to
produce the soft breaking parameters in many of his models through retrofitting, in
a way which naturally yields the required approximate R symmetries.

While these models have virtues for understanding CP violation,
they possess another potential difficulty.  With multiple messengers,
supersymmetry breaking can induce a non-zero $D$-term for hypercharge at one loop
(for a discussion, see \cite{dinefischler,dineminnesota}).  Additional special features are required
to suppress these.  Typically one invokes a ``Messenger parity" symmetry.
By itself, this symmetry is necessarily violated by the gauge couplings
of the MSSM fields, but such an approximate symmetry can arise
as an accident of other symmetries, as will be discussed
elsewhere.  In addition,
from our earlier discussion, it is clear that $R$ symmetry, by itself, is not
enough to guarantee suppression of CP violation.  If there are several singlets, $X_i$, of $R$
charge $2$, then the unified formula for the gaugino masses does not hold.

\subsubsection{Retrofitting Cheung et al's Model}

Rewriting the model above in the notation of \cite{shihnew}:
\begin{eqnarray}
W= \lambda X (\tilde{\phi}_1\phi_2 + \tilde{\phi}_2\phi_2) + m\tilde{\phi}_2\phi_1 + FX
\end{eqnarray}
Such a model has an R-symmetry. Following the analysis of \cite{shih},
the Coleman Weinberg potential for this model can lead to
$\langle X \rangle \not= 0$,
spontaneously breaking the R-symmetry and making such models phenomenologically viable.
In the case that $F$ and $m_i$ are generated dynamically from gaugino condensation we have a
discrete R-symmetry under which $R(F)=2$ and $R(m_i)=1$.
The Superpotential can be constrained to be of this form if,
for example, $R(W)=2$,$R(X)=-2$,
$R(\tilde{\phi}_1)=1$,$R(\phi_1)=3$,$R(\phi_2)=5$, and $R(\tilde{\phi}_2)=-1$. 
(and there are no $Z_N$ equivalences).


\subsection{R Symmetries:  Explicit Breaking}

Achieving explicit breaking in a model where all scales are comparable,
and in which the structure of the underlying lagrangian
is dictated by symmetries, turns out to be challenging; it is not
simple to achieve metastable supersymmetry
breaking in a natural way.  To illustrate
the problems, consider adding  messengers to the simple O'Raifeartaigh
model.
We can take these to be a $5$ and $\bar 5$ of $SU(5)$ ($q,\bar q,
\ell,\bar \ell$), with
couplings
\beq
W = Z \mu^2+ Z A^2 + m YA +(m_3 +  \lambda_3 Z) q  \bar q +(m_2 + \lambda_2 Z) \ell \bar \ell .
\label{mess}
\eeq
The terms $m_1,m_2$ as well as the supersymmetry breaking scale ($F_Z$)
can be generated dynamically. Classically, supersymmetry is broken
in this theory; provided $\vert m_i + \lambda_i Z \vert > \vert \mu \vert$,
the $q,\bar q,\ell,\bar \ell$ fields have positive mass-squared.
$\langle Z \rangle$ is undetermined classically.  At one loop, the
standard Coleman Weinberg computation gives a minimum, whose precise
value depends on the parameters in the superpotential.  At the minimum,
the fields $q,\bar q$ and $\ell,\bar \ell$ have supersymmetric masses,
\beq 
m_q = m_3 + \lambda_3 Z~~~m_\ell = m_2 + \lambda_2 Z.
\eeq 
Instead of the usual gauge-mediated formula, we now have the formula
of eqn. \ref{scalarsemgm}, with
\beq
\Lambda_q^2 = {\lambda_3^2 F_Z \over m_q}~~~~
\Lambda_\ell^2 = {\lambda_2^2 F_Z \over m_\ell}.
\eeq
So we have a spectrum in which the mass formula of MGM does not hold,
and in which compression of the spectrum is possible.

Not surprisingly, 
it is difficult to obtain this structure as a consequence
of symmetries, 
if $\mu$ and $m$ are to be the same order.  Examining equations
\ref{retro},\ref{ww}, one sees that
the $R$ charge of $Z$ and $m_i$ and $\mu$ must be the same.
This necessarily means that $Z^3$, $m Z^2$ and $m^2 Z$ (and others) are allowed
by the symmetry.  In order that there
be a metastable ground state within a distance of
order $m_i$ from the origin, the coefficients of each of these terms must be
extremely small,
substantially smaller than the loop factors which give
rise to the $Z$ potential near the origin.

On the other hand, we are used to the idea that there are very small
Yukawa couplings in nature.  A standard scenario for
generating such small couplings is provided by the
Froggatt-Nielsen mechanism.  Here there is some field, $\phi$ (for convenience
taken to be dimensionless), transforming
under additional symmetries, such that $\phi$ is a small number (in our example below,
we will need $\phi$ small compared to a loop factor).  Adopting this approach
will imply a hierarchy between the masses of the messengers and the
fundamental supersymmetry breaking scale.

Introducing such a field, we can consider a superpotential of the form:
\beq
W = m^2\phi^2 Z - Z A^2 + m \phi Y A + m\phi^2 \bar 5 5 + Z\phi^2 \bar 5 5
\label{messengerw}
\eeq
where we are being schematic in two ways:
\begin{enumerate}
\item  We are again not indicating explicitly dimensionless
numbers of order one.
\item  Couplings and masses such as $m \bar 5 5$ are shorthand for
$m_1 q \bar q + m_2 \ell \bar \ell$
\end{enumerate}
In this potential, the splittings of the scalar components of $5,\bar 5$ ($m^2 \phi^4$) are
comparable to the masses-squared of the fields, necessary if the messenger
masses are to be only a loop-factor removed from the masses of the squarks
and sleptons.  Note, however, that the messenger fields are lighter by a
factor of $\phi$ than the fields $A$ and $Y$.  
The lagrangian, classically, exhibits, for general values of $Z$,
supersymmetry breaking metastable ground states
 with $\langle 5 \rangle = \langle \bar 5 \rangle =0$; there is no approximate
 $R$ symmetry. 
$Z$ will be determined by quantum mechanical effects, which we will discuss shortly.

The superpotential of eqn. \ref{messengerw} is not the most general consistent
 with symmetries.
Under any would-be symmetry,
$Z$ transforms as $m$.  This means that $m \phi^2 Z^2$ and $\phi^2 Z^3$ are allowed,
as well as similar terms involving $Y$.  These additional terms are dangerous.
They lead to supersymmetric minima, and they yield tadpoles and mass corrections
which may destabilize the non-supersymmetric minima if $\phi$ is not sufficiently
small.  
To understand the constraints
on $\phi$, consider the Coleman-Weinberg calculation for the system
represented by \ref{messengerw}.  The largest contribution
to the potential comes from loops containing the field
$A$.  This calculation is standard, and gives a minimum
for $Z$ at the origin.    The curvature of the Coleman-Weinberg
potential is of order
\beq
V^{\prime \prime} \sim \alpha {\vert F_Z \vert^2 \over m^2 \phi^4}\sim \alpha {m^2}.
\eeq
Here $\alpha$ denotes a loop factor ($\lambda^2 \over 16 \pi^2$, where $\lambda$ is a typical
dimensionless coupling in the superpotential).
Including $5,\bar 5$ loops gives a small $Z$ tadpole
\beq
\delta V^{\prime} \sim \alpha m^2 \phi^{12}.
\eeq
These lead to tiny shifts in $Z$ (${\cal O}(\phi^6)$).

Now including the coupling $m \phi^2 Z^2$
gives, classically,
a tadpole for $Z$ of order $m^3 \phi^4$; this leads to a shift in $Z$
of order $\alpha^{-1} m \phi^2$.  However, the $Z$ potential varies (in field space) on
scales $m \phi$, so for $\phi < \alpha$ this is a small shift.  Similarly,
the $Z^3$ coupling generates a mass term for $Z$ near the origin of order $m^2 \phi^2$; this
is larger than the Coleman Weinberg contribution by a factor of order $\alpha^{-1}$.

The basic structure of the model is technically natural; renormalizable couplings
other than those we have mentioned can be forbidden by symmetries.  A sample
set of symmetries are listed in the table below:

\begin{center}
\begin{tabular}{|c|c|c|c|}
  \hline
  Field & R
  & $R^{\prime}$ & $Z_2$ \\
  \hline
 W & 3 &  7 & 0 \\
 m & 1 &  1 & 0  \\
  A & 1 & 3 &  -1 \\
 Y & 1 & 1 & -1 \\
 Z & 1 & 1 & 0\\
 $\phi$ & 0 &  2 & 0 \\
$5 \bar 5$ & 2 & 2 & 0 \\
  \hline
\end{tabular}
\end{center}

\vskip .5cm
\noindent
The first of these symmetries accounts for the absence of large mass terms.
The second explains the suppression by various powers of $\phi$.  The third
forbids certain potentially dangerous couplings, like $Y A^2$.

It is interesting that to an observer at the scale $m_i$,the superpotential does not appear generic,
and there is no relic
of the symmetries responsible for this at low energies.  The observer would simply
note that the structure is preserved in perturbation theory, due to 
non-renormalization theorems.

In these models, while the messenger scale can be of order
$100$ TeV, the underlying scale of supersymmetry breaking
is necessarily significantly larger.  The messenger scale is $\phi^2 m$; the
splittings among the messengers are of the same order, so this scale might be
as low as $100$ TeV or so.  So, along with compression of the spectrum,
these models can ameliorate the fine tuning problems
of gauge mediated models.  However, the susy-breaking $F$ term, which determines the
gravitino mass, is of order $(\phi m)^2$, and the gravitino mass correspondingly larger.
This is potentially problematic for dark matter\cite{riotto},
requiring a non-standard cosmology above the weak scale.

\subsection{CP Violation}

Models with {\it spontaneous} breaking of $R$ symmetry
often have the virtue that CP violation
can automatically be small or zero.  If, for example, $Z^0$
in eqn. \ref{dm} is replaced by its expectation
value, i.e. if fluctuations of $Z$ are ignored, then through field redefinitions,
all of the couplings of the lagrangian may be made real; CP is, at the very least,
loop suppressed.  As we will shortly
see, this feature can be preserved in some cases by the dynamics
which generates the supersymmetric Higgs mass, $\mu$\cite{giudice,giudicetwo}.  
Cheung et al\cite{shihnew} have
shown that in a broad class of models with $R$ symmetries
at low energies, even when there are CP violating phases, gaugino
masses all carry the same phase, and one loop contributions to
electric dipole moments vanish.  But we have also seen that $R$ symmetries
are not, by themselves, enough.

If the breaking of the symmetry
is explicit, CP violation is more problematic.    
First, it should be noted that even if this framework is embedded in a conventional
GUT, one does not expect GUT relations to hold for the messenger mass terms.  These
arise from couplings, for example, to $W_\alpha^2$, and, in general, these will be
one loop effects in the underlying theory.  Alternatively, these couplings may arise
from instantons, and there is, in general, no reason to expect that they should
obey GUT relations. 

In theories without automatic suppression of CP violation, we need to suppose
that CP-violation is inherently small.  This can arise if CP is spontaneously
broken by a small amount.  This still permits an order one KM phase, while allowing
suppression of CP-violating phases in the messenger and supersymmetry violating
phases, much along lines suggested some time ago by Nir and Rattazzi\cite{nirrat}

\subsection{$\mu$, $B_\mu$ and $A$}

Various proposals have been put forth for generating $\mu$ in the framework of gauge mediation.
Many of these involve additional degrees of freedom, beyond those of the MSSM.
One of the most attractive approaches is that of \cite{giudice,giudicetwo}.   Here
one considers two sets of messenger fields, $5_1,5_2$ and $\bar 5_1,\bar 5_2$.
One has couplings
\beq
Z (5_1 \bar 5_1 + 5_2 \bar 5_2) + S 5_1 \bar 5_2 + S H_U H_D +S^3.
\label{giudicemu}
\eeq
One can readily check that the one loop contribution to the mass of $S$ vanishes,
but the two loop contribution is negative, for suitable choice of parameters.
As a result, $S$ obtains a vev of one loop order, while $F_S$ is two loop order.
As a result, $B_\mu \sim \mu^2$.
In the context
of models with spontaneous breaking of the $R$ symmetries, such an approach was followed
in \cite{dinemason};  in both this framework
and the framework of explicit $R$ symmetry breaking, one can obtain
models whose structure is enforced by symmetries.
As in all multimessenger models, one requires some sort of messenger parity
symmetry to suppress one loop D-term contributions to sfermion masses.  In eqn.
\ref{giudicemu}, this can be an accidental consequence of a discrete symmetry interchanging
the fields $5_1,\bar 5_1$ with $5_2,\bar 5_2$.
As discussed above, with
spontaneous breaking, one can sometimes understand the suppression of CP violating
effects.  This is not true with explicit violation, as we have already seen even
before including the Higgs fields.

An alternative is to generate 
$\mu$ by ``retrofitting", just as we did the relevant
parameters of the O'Raifeartaigh models.  We could include
a term
\beq
\gamma{W_\alpha^2 \over M^2}
 H_U H_D
 \eeq
in the superpotential.  The structure, again, can be enforced by
$R$ symmetries.    It is necessary that $\gamma$ be small, of order
a loop factor.  We can simply postulate a small number, or suppose that
this is small from a FN mechanism, just as we accounted for small,
dimensionless numbers in the previous sections.  
Couplings of $X$ or $Z^0$ can be forbidden by the same $R$ symmetries,
so classically, there is no $B_\mu$ term.  Instead, $B_\mu$ is generated
by one loop effects, so $\tan(\beta)$ is large.

In the case of explicit breaking, one can again generate $\mu$ by
retrofitting, writing, e.g.
\beq
\phi^3 {W_\alpha^2 \over M^2}
 H_U H_D
 \eeq
However, since $Z$ transforms under any symmetry like $m$, one cannot forbid
a coupling $Z \phi^3 H_U H_D$, so
\beq
\vert B_\mu \vert^2 \gg \vert \mu \vert^2.
\eeq
Similar problems arise if we try to couple $A$ or $Y$ to $H_U H_D$, and even
if we allow hierarchies involving powers of $\phi$ between different scales (this
can easily be shown quite generally).

With additional fields, we can account for $\mu$ in a natural way.  As a simple example,
we introduce fields $X$ and $B$, with superpotential:
\beq
W_H = X(B^2 - \phi^4 m^2) + \phi B H_U H_D
\eeq
This leads to $\mu \sim m \phi^3$, which is the desired order of magnitude.  Since $B$
has no $F$ component, classically, $B_\mu$ vanishes.  We can
banish couplings of $Z$ to $H_U H_D$ by symmetries  Under our symmetries, we can assign charges
as follows:

\begin{center}
\begin{tabular}{|c|c|c|c|}
  \hline
  Field & R
  & $R^{\prime}$ & $Z_2$ \\
  \hline
 $X$ & 1 &  -3 & 0 \\
 $B$ & 1 &  5 & -1  \\
 $H_U H_D$ & 2 &  0 & -1 \\
  \hline
\end{tabular}
\end{center}

\vskip .5cm

This construction avoids the need for messenger parity.  But,
as before, CP is an issue in this framework.  Again, one possible solution is that
the field(s) $\phi$ spontaneously break $CP$.  This can suppress the phases
of gaugino masses.  The phases associated with $B$, $X$ and the Higgs fields
can then be absorbed in field redefinitions.


If we assume that this is the
origin of $\mu$, we can ask about the origin of $B_\mu$.  In the low energy theory, there is 
a one loop diagram which contributes to an $H_U H_D$ term in the potential, containing an
internal wino and Higgsino.  This gives an $H_U H_D$ mass term of order 
\beq
B_\mu \approx {\alpha_w \over 2 \pi}\mu m_w \vert \ln(\phi) \vert.
\eeq 
This leads, in turn, to a value of $\tan(\beta)$ of order 
\beq
\tan \beta \sim {2 \pi \over \alpha_w \vert  \ln(\phi) \vert},
\eeq
which, depending on the precise values of the masses, may or may not be acceptable.

Let us return briefly to models with spontaneous breaking of the $R$ symmetry and
ask about alternative mechanisms for generating the $\mu$ and $B_\mu$ terms
through retrofitting.  Coupling a susy-breaking field, $X$, to $H_U H_D$ poses
the usual $\mu-B_\mu$ problem; generating a small coupling by
introducing a field $\phi$, by itself, does not help.  So it is simplest to introduce,
again, additional fields analogous to $B$ and $X$.

\section{Conclusions}

A priori, in speculating about the possibility of gauge-mediated supersymmetry
breaking, one can contemplate phenomena over many decades of energy.  We have motivated
a possible role for very low energy gauge mediation by considerations of fine
tuning.  It is well-known, however, that the phenomenological possibilities of such
low energy breaking are potentially quite interesting, including dramatic
decay channels and signatures.

We have suggested that such low energy breaking arises most naturally in the framework
of O'Raifeartaigh models whose dimensionful parameters all arise from non-perturbative
effects in a theory which is weakly coupled at high energies.
Particularly important is the question of $R$ symmetry
breaking.  In models with approximate $R$ symmetries (spontaneously
broken), one can contemplate a compressed spectrum,
approximate CP conservation, and interesting mechanisms
for generating $\mu$ and $B_\mu$, in a natural framework (i.e.
structures enforced by symmetry).  For models with explicit breaking, many
of these same features are readily obtained, though understanding CP violation is more
challenging, and a hierarchy between the messenger scale and the underlying
supersymmetry-breaking scale at least of order a loop factor seems required.

From a low energy point of
view, these are simply O'Raifeartaigh models.  In
the high scale breaking models, 
what is surprising to a low energy observer
is that the models do not appear natural (generic); one needs a microscopic understanding
to see that they are the most general models consistent with symmetries.  Indeed, this
picture is a realization of a number of Witten's original vision for dynamical supersymmetry
breaking\cite{wittendsb}:
\begin{enumerate}
\item  There are small parameters, as a consequence of short distance non-perturbative
effects
\item  There are couplings which vanish, for no apparent reason, in the low energy
theory.
\item  The consistency of the first two points is a consequence of the non-renormalization
theorems for the couplings of the low energy theory.  In the case of hierarchical breaking,
the vanishing of certain superpotential terms can be understood in terms of the symmetries
of the microscopic theory.
\end{enumerate}
 
Surely more elegant models can be constructed than those in this paper, which
are presented mainly to provide an existence proof of simple and sensible models with very low
energy supersymmetry breaking.

\noindent
{\bf Acknowledgements:}
We thank N. Arkani-Hamed and Scott Thomas for many discussions of the little
hierarchy and the problems of CP violation, and Nathan Seiberg
for discussions of the classification of supersymmetric theories.
We thank David Shih for discussions of his recent work, and Linda Carpenter
for general discussions of issues in gauge mediation.
This work supported in part by the U.S.
Department of Energy and the J.S. Guggenheim Foundation.  Part of this work was done at
the Institute for Advanced Study; M.D. thanks the IAS
for its hospitality and support.


\begin{thebibliography}{9}



\bibitem{wittendsb}
  E.~Witten,
  Nucl.\ Phys.\  B {\bf 188}, 513 (1981).

\bibitem{iss}
  K.~Intriligator, N.~Seiberg and D.~Shih,
  JHEP {\bf 0604}, 021 (2006)
  [arXiv:hep-th/0602239].

\bibitem{dfs}
  M.~Dine, J.~L.~Feng and E.~Silverstein,
  arXiv:hep-th/0608159.



\bibitem{shih}
  D.~Shih,
  arXiv:hep-th/0703196.


\bibitem{dinemason}
  M.~Dine and J.~Mason,
  arXiv:hep-ph/0611312.


\bibitem{aharonykachrusilverstein}
  O.~Aharony, S.~Kachru and E.~Silverstein,
  arXiv:0708.0493 [hep-th].

\bibitem{seibergreview}
  K.~Intriligator and N.~Seiberg,
  arXiv:hep-ph/0702069.

\bibitem{shihnew}
  C.~Cheung, A.~L.~Fitzpatrick and D.~Shih,
  arXiv:0710.3585 [hep-ph].


\bibitem{bankskorneel}
T. Banks and Korneel van den Broek, to appear.

\bibitem{pentagon}
  T.~Banks,
  arXiv:hep-ph/0510159;
  T.~Banks,
  arXiv:hep-ph/0606313.

\bibitem{iss2}
  K.~Intriligator, N.~Seiberg and D.~Shih,
  JHEP {\bf 0707}, 017 (2007)
  [arXiv:hep-th/0703281].

\bibitem{giudice}
  G.~F.~Giudice and R.~Rattazzi,
  Nucl.\ Phys.\  B {\bf 511}, 25 (1998)
  [arXiv:hep-ph/9706540].


\bibitem{giudicetwo}
  A.~Delgado, G.~F.~Giudice and P.~Slavich,
  Phys.\ Lett.\  B {\bf 653}, 424 (2007)
  [arXiv:0706.3873 [hep-ph]].

\bibitem{agashegraesser}
  K.~Agashe and M.~Graesser,
  Nucl.\ Phys.\  B {\bf 507}, 3 (1997)
  [arXiv:hep-ph/9704206].


\bibitem{dineminnesota}
  M.~Dine,
  Nucl.\ Phys.\ Proc.\ Suppl.\  {\bf 62}, 276 (1998)
  [arXiv:hep-ph/9707413].



\bibitem{focuspoint}  For two examples, see
  J.~L.~Feng and K.~T.~Matchev,
  Phys.\ Rev.\  D {\bf 63}, 095003 (2001)
  [arXiv:hep-ph/0011356];
  R.~Essig and J.~F.~Fortin,
  arXiv:0709.0980 [hep-ph].

\bibitem{bmssm}
  A.~Brignole, J.~A.~Casas, J.~R.~Espinosa and I.~Navarro,
  Nucl.\ Phys.\  B {\bf 666} (2003) 105
  [arXiv:hep-ph/0301121];
  J.~A.~Casas, J.~R.~Espinosa and I.~Hidalgo,
  JHEP {\bf 0401}, 008 (2004)
  [arXiv:hep-ph/0310137];
  M.~Dine, N.~Seiberg and S.~Thomas,
  arXiv:0707.0005 [hep-ph].

\bibitem{dinefischler}
  M.~Dine and W.~Fischler,
  Phys.\ Lett.\  B {\bf 110}, 227 (1982).

\bibitem{murayamaiss}
  H.~Murayama and Y.~Nomura,
  Phys.\ Rev.\  D {\bf 75}, 095011 (2007)
  [arXiv:hep-ph/0701231].


\bibitem{aharonyseiberg}
  O.~Aharony and N.~Seiberg,
  JHEP {\bf 0702}, 054 (2007)
  [arXiv:hep-ph/0612308].


\bibitem{itiy}
  K.~I.~Izawa and T.~Yanagida,
  Prog.\ Theor.\ Phys.\  {\bf 95}, 829 (1996)
  [arXiv:hep-th/9602180];
  K.~A.~Intriligator and S.~D.~Thomas,
  Nucl.\ Phys.\  B {\bf 473}, 121 (1996)
  [arXiv:hep-th/9603158].





\bibitem{kitano}
  M.~Ibe and R.~Kitano,
  arXiv:0711.0416 [hep-ph].


\bibitem{dinenelson}
See, e.g.  M.~Dine, A.~E.~Nelson, Y.~Nir and Y.~Shirman,
  Phys.\ Rev.\  D {\bf 53}, 2658 (1996)
  [arXiv:hep-ph/9507378].



\bibitem{murayamaitiy}
  H.~Murayama,
  Phys.\ Rev.\ Lett.\  {\bf 79}, 18 (1997)
  [arXiv:hep-ph/9705271].

\bibitem{riotto}
  M.~Viel, J.~Lesgourgues, M.~G.~Haehnelt, S.~Matarrese and A.~Riotto,
  Phys.\ Rev.\  D {\bf 71}, 063534 (2005)
  [arXiv:astro-ph/0501562].

\bibitem{nirrat}
  Y.~Nir and R.~Rattazzi,
  Phys.\ Lett.\  B {\bf 382}, 363 (1996)
  [arXiv:hep-ph/9603233].








\end{thebibliography}
\end{document}